\documentclass[12pt]{amsart}
\usepackage{amsmath}  

\newtheorem{lemma}{Lemma}
\newtheorem{proposition}{Proposition}
\newtheorem{theorem}{Theorem}
\newtheorem{remark}{Remark}

\newtheorem{corollary}{Corollary}
\newcommand{\Z}{\mathbb Z}
\newcommand{\Zp}{\mathbb Z^+}
\newcommand{\N}{\mathbb N}
\newcommand{\C}{\mathbb C}
\newcommand{\K}{\mathcal K}
\newcommand{\bo}{\mathcal B([0,2\pi))}
\newcommand{\torus}{\mathbb T}
\newcommand{\boto}{\mathcal B(\torus)}
\newcommand{\lto}{L_2(\torus,\mu)}
\newcommand{\ltok}{L_2(\torus\times\torus,\mu\times\mu)}
\newcommand{\fii}{\varphi}
\newcommand{\hi}{\mathcal H}
\begin{document}

\sloppy

\title[Covariant localizations in  torus and  phase observables]{Covariant localizations in  
the torus and  the phase observables}

\author{G. Cassinelli}
\address{Gianni Cassinelli, Dipartimento di Fisica,
Universit\`a di Genova, I.N.F.N., Sezione di Genova, Via Dodecaneso~33,
16146 Genova, Italy}
\email{cassinelli@ge.infn.it}
\author{E. De Vito}
\address{Ernesto De Vito, Dipartimento di Matematica, Universit\`a di
Modena, Via Campi 213/B, 41100 Modena, Italy and I.N.F.N., Sezione di Genova,
Via Dodecaneso~33, 16146 Genova, Italy}
\email{devito@unimo.it}
\author{P. Lahti} 
\address{Pekka Lahti, Department of Physics, University of Turku,
20014 Turku, Finland}
\email{pekka.lahti@utu.fi}
\author{J.-P. Pellonp\"a\"a} 
\address{Juha-Pekka Pellonp\"a\"a, Department of Physics, University of Turku,
20014 Turku, Finland}
\email{juhpello@utu.fi}
\date{\today}
\begin{abstract}
We describe all the localization observables of a quantum particle in a one-dimensional box
in terms of sequences of unit vectors in a Hilbert space.
An alternative representation  in terms of positive semidefinite complex matrices is 
furnished and  the commutative localizations are singled out.   As a consequence, we also get 
a vector sequence characterization of the covariant phase observables. 
\end{abstract}
\maketitle

\section{Introduction}

We investigate the problem of the localization of a free quantum particle
moving in a one-dimensional box with periodic boundary conditions, adopting the
point of view that observables are represented as appropriate normalised positive operator measures.
(For that approach, see, e.g. \cite{Davies, Holevo, Ludwig, OQP}).
Therefore,  if one chooses  the one-dimensional torus $\torus$ as the configuration space of the system, then  a
localization observable $E$ is a map that defines for any (Borel) subset $X\subset\torus$ 
a bounded operator $E(X)$ such that,  
if $\psi$ is the (vector) state of the system, 
the number $\langle \psi|E(X)\psi\rangle$ is the probability that a localization  measurement 
of the particle in that state  leads to a result in $X$.
%
%
The group of motions of the system is the torus itself that acts on the vector states by
means of the {\em geometric} representation. 
The basic requirement for $E$ to represent localization observable is, therefore,
that $E$ is covariant with respect to this action.  
Hence  a localization observable is a normalized positive operator measure covariant under the
geometric action of the torus $\torus$. 

In the following we call such observables $\torus$-covariant localization
observables and we show that they are
characterized in terms of sequences of unit vectors in an infinite dimensional
Hilbert space. 
In this framework we select the measures that are projection
valued or commutative, and we discuss the problem of the equivalence of such
operator measures. As a by product, we also get a representation of the phase
observables, that is, the normalized positive operator measures which are
covariant under the shifts generated by the number observable.  Our proof is
based on a direct application of a theorem due to Cattaneo \cite{Cattaneo},
which generalizes Mackey's imprimitivity theorem for positive operator
measures. Instead, one could use the results of Holevo \cite{Holevo83,
  Holevo87}, based also on group theoretical arguments, to obtain a
classification in terms of measurable fields of sesquilinear forms, which in
the present context can be described as infinite dimensional positive
semidefinite complex matrices with diagonal elements equal one.  For sake of
comparison, we also derive the matrix characterization by direct methods,
using only basic analysis and measure theory. This approach has been used in
\cite{JPPL99} to work out all the phase observables in terms of {\em phase
  matrices}.

\section{ $\torus$-Covariant localization observables}\label{covloc}
Let $\torus =\{z\in\C\,:\, |z|=1\}$ denote the one dimensional torus,
regarded as a compact (second countable) Abelian group.
Let $\boto$ be
the Borel $\sigma$-algebra of  $\torus$,  $\mu$  the Haar measure
on $\torus$, $\lto$ the Hilbert space of  square integrable  Borel
functions $f:\torus\to\C$ and $\mathcal L(\lto)$  the set of bounded operators on $\lto$.

The group $\torus$ acts on $\lto$ via the geometric action
$$
[U(a)f](z)=f(az),\ \ \ a\in\torus,\  f\in\lto,\  z\in\torus,
$$
which is unitary and continuous with respect to the strong operator
topology.

A $\torus$-covariant localization observable is a  positive normalized
operator measure on $\torus$, $E:\boto\to\mathcal L(\lto)$,
such that, for all $X\in\boto, a\in\torus$,
\begin{equation}\label{covariant}
U(a)E(X)U(a)^* = E(aX).
\end{equation}
Since the action of $\torus$ on itself is transitive, Eq.~(\ref{covariant}) means that
$(U,E)$ is a transitive system of $\torus$-covariance
based on 
$\torus$ and, hence, $(U,E)$ is described by \cite[Proposition~2]{Cattaneo}.

In order to apply the cited result,  let us notice the following facts.  
The stability subgroup of any point of $\torus$ is the trivial
subgroup $\{1\}$.  The trivial representation $\sigma$ of
$\{1\}$ acting on  $\lto$ contains all the (trivial)
representations of $\{1\}$ and the corresponding imprimitivity system $(R,P)$  for
$\torus$ based on $\torus$ 
induced by $\sigma$  acts on  $\ltok\simeq L^2(\torus,\mu,\lto)$ as 
\begin{eqnarray*}
(R(a)\fii)(z_1,z_2) & = & \fii(az_1,z_2),\\
(P(X)\fii)(z_1,z_2) & = & \chi_X(z_1)\fii(z_1,z_2),
\end{eqnarray*}
where $\fii\in\ltok, a\in\torus, X\in\boto, z_1,z_2\in\torus$.

Proposition 2 of \cite{Cattaneo} shows that, given a
$\torus$-covariant localization observable $E$, there exists  
an isometry  
$$V:\lto\to\ltok,$$ 
which intertwines the action $U$ with $R$ and such that
\begin{equation}\label{correspondence}
E(X)=V^*P(X)V,\ \ \ \ X\in\boto.
\end{equation}
Conversely, given an intertwining isometry $V$ from $\lto$ to $\ltok$,
Eq.~(\ref{correspondence}) defines a  positive normalized operator measure $E$
satisfying Eq.~(\ref{covariant}).

Hence, to classify all the $\torus$-covariant localization
observables, one has to  determine all the isometric  mappings $V$ such that
\begin{equation}\label{intertwine}
VU(a)=R(a)V,\ \ \ a\in\torus.
\end{equation}

To perform this task, observe that  the monomials  $e_n$, $n\in\Z$,
$e_n(z)=z^n$, $z\in\torus$, form an orthonormal basis of $\lto$ and
the action of $U$ on them is diagonal, that is, 
$$
U(a)e_n = a^ne_n.
$$
Moreover, the vectors
$$
(e_n e_j)(z_1,z_2)=e_n(z_1)e_j(z_2)=z_1^n z_2^j,
$$ 
where $n,j\in\Z$,
form an orthonormal basis of $\ltok$, and the action of $R$ on them is simply
$$
R(a)e_n e_j= a^ne_n e_j.
$$
It follows that, for any 
$n\in\Z$, the subspace of $\lto$ generated by $e_n$ and  the subspace of $\ltok$ generated by $\{e_ne_j\}_{j\in\Z}$ 
carry the representation of $\torus$, $z\mapsto z^n.$  

Hence, if  $V:\lto\to\ltok$  is an isometry intertwining $U$ 
and $R$, for any $n\in\Z$, $Ve_n$ must be in the vector space 
$\overline{\mathrm{span}}\{(e_ne_j)\}_{j\in\Z}\simeq\lto$, that is, 
$Ve_n=e_n h_n$ for some unit vector $h_n$ in $\lto$.

Conversely, if $(h_n)_{n\in\Z}$ is a sequence of unit vectors in $\lto$, then the
mapping $e_n\mapsto e_nh_n$ extends to a unique linear isometry $V:\lto\to\ltok$
which intertwines the actions $U$ and $R$ and, by means of Eq.~(\ref{correspondence}), 
the corresponding operator measure $E$ is explicitly given by, 
\begin{eqnarray*}
\langle e_n|E(X)e_m\rangle
&=&
\langle e_n|V^*P(X)Ve_m\rangle\\
&=&
\langle Ve_n|P(X)Ve_m\rangle\\
&=&
\langle e_nh_n|P(X)e_mh_m\rangle\\
&=&
\langle h_n|h_m\rangle\,\int_Xz^{m-n}\,\mathrm d\mu(z),
\end{eqnarray*}
where $n,m\in\Z$. Then, if $|e_n\rangle\langle e_m|$ denotes the rank one operator
$\lto\ni f\mapsto \langle e_m|f\rangle\,e_n\in\lto$ 
we may thus write, for all $X\in\boto$,
\begin{equation}\label{locobs}
E(X) =\sum_{n,m\in\Z}\langle h_n|h_m\rangle\, \int_X z^{m-n}\,\mathrm d\mu(z)\, |e_n\rangle\langle e_m|,
\end{equation}
where the double series converges in the weak operator topology.
We observe  that two sequences of unit vectors $(h_n)_{n\in\Z}$ and $(k_n)_{n\in\Z}$ define the
same $\torus$-covariant localization observable if and only if
$\langle h_n|h_m\rangle=\langle k_n|k_m\rangle$ for all $n,m\in\mathbb Z$.

For sake of completeness we also compute  the adjoint map $V^*:\ltok\to\lto$.
We get
\begin{eqnarray}
\langle V^*e_n e_j|e_p\rangle
&=& \langle e_n e_j|Ve_p\rangle\nonumber\\
&=& \langle e_n e_j |e_p h_p \rangle\nonumber\\
&=&\left\{ \begin{array}{ll}
0, & \textrm{when $n\ne p$,}\\
  \langle e_j | h_n \rangle, &\textrm{when $n=p,$}
\end{array}\right.
\end{eqnarray}
showing that for any $n,j\in\Z$,
\begin{equation}
V^*e_n e_j  = \langle h_n | e_j \rangle\, e_n.
\end{equation}

We now discuss the problem of the equivalence.  Two $\torus$-covariant
localization observables $E$ and $E'$ are equivalent if there is a
unitary operator
$W:\lto\to\lto$ such that, for all $a\in\torus$, $X\in\boto$,
\begin{eqnarray*}
WU(a) &=& U(a)W\\
WE(X)&=& E'(X)W,
\end{eqnarray*}
Clearly, this definition is the requirement that $(U,E)$ and $(U,E')$ are equivalent
as $\torus$-covariant systems.

The first condition implies now that  for each $n\in\Z$,  $We_n=z_ne_n$, for some $z_n\in\torus$.
Therefore, the equivalence of $(U,E)$ and $(U,E')$  equals the fact that
for each $n,m\in\Z$ and $X\in\boto$,
$$
\langle e_n|E'(X)e_m\rangle = z_n\overline z_m\langle e_n|E(X)e_m\rangle,
$$
for some $z_n,z_m\in\torus$. Then, taking into account Eq.~(\ref{locobs}), two sequences of unit vectors
$(h'_n)_{n\in\Z}$ and $(h_n)_{n\in\Z}$ define equivalent
$\torus$-covariant localization observables if and only if for each $n,m\in\Z$,
$$
\langle h'_n|h'_m\rangle = \langle z_nh_n|z_mh_m\rangle.
$$

Finally, we consider the problem of projection valued measures. 
Let  $\K$ be the closed subspace of $\ltok$ generated by $\{P(X)e_nh_n\, : \, X\in\boto, n\in\Z\}$. 
The space $\K$ is stable under the action of the imprimitivity system $(R,P)$. Since the projection 
measure $P$ acts only on the vector $e_n$ and $\{\chi_Xe_n\, : \, X\in\boto\}$ generates $\lto$, one has 
$$\K\simeq\lto\otimes L,$$
where $L$ is the closed subspace generated by the vectors $h_n, n\in \Z$.
According to \cite[Proposition~1 ]{Cattaneo},
$E$ is projection valued if and only if $V(\lto)=\K$, that is, $L$ is one dimensional.

We summarize the above construction in form of a theorem.

\begin{theorem}\label{localization}
Any $\torus$-covariant localization observable $E:\boto\to\lto$ is of the form
$$
E(X) =\sum_{n,m\in\Z}\langle h_n|h_m\rangle\, \int_X z^{m-n}\,\mathrm d\mu(z)\, |e_n\rangle\langle e_m|,
\ X\in\boto,
$$
for some sequence of unit vectors $(h_n)_{n\in\Z}$ in $\lto$. 
Two sequences of unit vectors $(h_n)_{n\in\Z}$  and $(k_n)_{n\in\Z}$ in $\lto$ determine
the same $\torus$-covariant localization observable  if and only if
$\langle h_n|h_m\rangle =\langle k_n|k_m\rangle$ for all $n,m\in\Z$.

Two such operator measures $E$ and $E'$ are equivalent
if and only if 
$$\langle h'_n|h'_m\rangle = \langle z_nh_n|z_mh_m\rangle,$$
for some sequence $(z_n)_{n\in\Z}$ in $\torus$.

The operator measure $E$ is projection 
valued exactly when the vectors $h_n$, $n\in\Z$, are of the form $h_n=z_nh$ for some unit vector $h$
and phase factors $z_n\in\torus$.
\end{theorem}

\subsection{Commutative localizations}\label{covloc3}
By means of the above theorem, we are now in position to characterize the commutative 
$\torus$-covariant localization observables. We recall that such an observable $E$ is commutative if  
$E(X)E(Y)=E(Y)E(X),$ for all $X,\,Y\in\boto$, that is, if $E$ is a
commutative operator measure.

Let $(h_n)_{n\in\Z}$ be a sequence of unit  vectors in $\lto$, $E$ 
the corresponding operator measure given by Theorem~\ref{localization}, 
and define, for all $n,m\in\Z$,  $c_{n,m}=\langle h_n | h_m
\rangle$. 
\begin{proposition}\label{commutative}
The $\torus$-covariant localization observable $E$  is commutative if and only if 
\begin{equation}\label{nmk}
c_{n,n+k}c_{n+k,m}=c_{n,m-k}c_{m-k,m}
\end{equation}
for all $n,\,m,\,k\in\Z$.
\end{proposition}

\begin{proof}
Define $\mu_{n,m,Y}(X):=\langle n|[E(X)E(Y)-E(Y)E(X)]|m\rangle$ for all
$n,\,m\in\Z$ and $X,\,Y\in\boto$.
Let $k\in\Z$, and calculate
$$
\int_\torus z^k\mathrm d\mu_{n,m,Y}(z)=
[c_{n,n+k}c_{n+k,m}-c_{n,m-k}c_{m-k,m}]\int_Yz^{n+k-m}
\mathrm d\mu(z).
$$
If $E(X)E(Y)=E(Y)E(X)$ for all $X,\,Y\in\boto$ then $\mu_{n,m,Y}(X)=0$ and, thus,
$c_{n,n+k}c_{n+k,m}=c_{n,m-k}c_{m-k,m}$ for all $n,\,m,\,k\in\Z$.

Conversely, if $c_{n,n+k}c_{n+k,m}=c_{n,m-k}c_{m-k,m}$, $n,\,m,\,k\in\Z$,
holds then 
\begin{eqnarray*}
\mu_{n,m,Y}(X)&=&\sum_{k=-\infty}^\infty(c_{n,n+k}c_{n+k,m}-c_{n,m-k}c_{m-k,m})\\
&&\times
\int_{X}z^{-k}\mathrm d\mu(z)
\int_{Y}z^{k+n-m}\mathrm d\mu(z)=0
\end{eqnarray*}
for all $n,\,m\in\Z$ and $X,\,Y\in\boto$. Therefore, 
$E(X)E(Y)=E(Y)E(X)$ for all $X,\,Y\in\boto$.
\end{proof}

An example of commutative $\torus$-covariant localization observable
is the following one. Let $\xi\in[-1,1]$ and $\phi,\psi\in\lto$ two unit vectors such that 
$$\langle\psi|\phi\rangle =\xi. $$
Consider the  sequence of unit vectors $( h_n)_{n\in\Z}$, 
with
\begin{eqnarray*}
&& h_n=\psi,\ {\rm  for\ even }\  n\  {\rm (including\  0)},\\
&&  h_n=\phi,\ {\rm  for\  odd }\ n.
\end{eqnarray*}
The  coefficients  $c_{n,m}=\langle h_n| h_m\rangle$,  $n,m\in\Z$, satisfy 
condition  (\ref{nmk})  so that   the corresponding $\torus$-covariant localization observable 
$E^\xi$ is commutative. Notice that $E^\xi$ is projection valued  if and only if $\xi=\pm1$.

\subsection{Matrix  characterization}\label{covloc4}
To end this section, we discuss an alternative characterization of the $\torus$-covariant
localization observables. 
If follows from Theorem~\ref{localization} that the operator measure
$E$ is uniquely defined in terms of the complex matrix 
elements $c_{n,m}=\langle h_n| h_m\rangle$, $n,m\in\Z$. It is clear that they satisfy the following two conditions:
\begin{itemize}
\item[{\rm (a)}] $c_{n,n}=1$, for all $n\in\Z$,
\item[{\rm (b)}] $\sum_{n,m=-k}^k c_{n,m}|e_n\rangle\langle e_m|\ge O$, for all $k\in\N$. 
\end{itemize}
Conversely, it is known, see, for example,  \cite[Chpt. 3]{Bergetal}, that given a family of complex numbers 
$\{c_{n,m}\in\C\,|\,n,m\in\Z\}$ which has the properties (a) and (b),
there exists a sequence of unit vectors $(h_n)_{n\in\Z}$  such that 
$c_{n,m}=\langle h_n| h_m\rangle$ and, hence, a $\torus$-covariant
localization observable $E$ defined by
$$
E(X)=\sum_{n,m\in\Z} c_{n,m} \int_X
z^{m-n}\mathrm d\mu(z)|e_n\rangle\langle e_m|,
$$
for all  $X\in\boto$.

For completeness, we give a simple construction of
a sequence of unit vectors which generates the matrix. The construction is 
slightly more general than actually needed here.

Let $J\subseteq\mathbb Z$ (especially $J=\mathbb Z$ or $J=\mathbb N$).
A  matrix $(b_{n,m})_{n,m\in J}$ is {\em positive semidefinite} if
for all sequences $(d_n)_{n\in J}\subset\mathbb C$, 
for which $d_n\ne0$ for only finitely many $n\in J$, 
$$
\sum_{n,m\in J}\overline{d_n}b_{n,m}d_m\ge0.
$$
For $J=\Z$ this is  equivalent to the above condition (b).   
(Condition (a) is equivalent to the fact that $\|h_n\|=1$ for all $n\in \Z$).
 
\begin{proposition}\label{jono}
Fix $J\subseteq\mathbb Z$. Let $\ell_2(J)$ be a sequence space
with the basis $(\chi_{\{n\}})_{n\in J}$.
A  matrix $(b_{n,m})_{n,m\in J}$ is positive semidefinite if and only if
there is a 
sequence $( h_n)_{n\in J}$ of vectors of $\ell_2(J)$ such that
$b_{n,m}=\langle h_n| h_m\rangle$ for all $n,\,m\in J$.
\end{proposition}

\begin{proof}
Consider a sequence $( h_n)_{n\in J}$ of vectors of $\ell_2(J)$
and put  $b_{n,m}=\langle h_n| h_m\rangle$.
If $(d_n)_{n\in J}\subset\mathbb C$ is a sequence 
for which $d_n\ne0$ for only finitely many $n\in J$, then
$$
\sum_{n,m\in J} \overline{d_n} b_{n,m}d_m =
\left\langle\left(\sum_{n\in J}d_n h_n\right)\bigg|\left(\sum_{m\in J}d_m h_m\right)\right\rangle \ge O,
$$
the sums being finite.

Suppose then that $(b_{n,m})_{n,m\in J}$ is positive semidefinite.
It follows that $b_{n,n}\ge0$, $b_{n,m}=\overline{b_{m,n}}$, and
\begin{equation}\label{1a}
\left|
\begin{array}{cc}
b_{n,n} & b_{n,m} \\
b_{m,n} & b_{m,m}  
\end{array} \right|=b_{n,n}b_{m,m}-|b_{n,m}|^2\ge 0
\end{equation}
for all $n<m$. Especially, if $b_{n,n}b_{m,m}=0$, then $b_{n,m}=0$.
Then the 
doubles series
$$
\sum_{n,m\in J\atop b_{n,n}\ne 0\ne b_{m,m}}\frac{b_{n,m}}{
\sqrt{b_{n,n}b_{m,m}}(|n|+1)(|m|+1)}|\chi_{\{n\}}\rangle\langle\chi_{\{m\}}|
$$
converges in the weak operator topology to a bounded and positive operator $S$.
Let $A$ be its square root 
and, for all $n\in J$,
$$ h_n:=\sqrt{b_{n,n}}(|n|+1)A\chi_{\{n\}}.$$
Then, taking into account that  $S=A^2$, one gets, for all $n,\,m\in J$
such that $b_{n,n}b_{m,m}\neq 0$, 
\begin{eqnarray*}
\langle h_n| h_m\rangle & = &\sqrt{b_{n,n}}(|n|+1)\sqrt{b_{m,m}}(|m|+1)
\langle \chi_{\{n\}}|A^2\chi_{\{m\}}\rangle \\
& = & b_{n,m}.
\end{eqnarray*}
If $b_{n,n}b_{m,m}= 0$, for example $b_{n,n}=0$, then $ h_n=0$ and, for
all $m\in J$, $b_{n,m}=0=\langle h_n| h_m\rangle$.
\end{proof}

The above proposition, when applied together with the natural isomorphism
$\ell_2(\Z)\ni\chi_{\{n\}}\mapsto e_n\in\lto $, gives then a vector sequence representation 
of the matrix $(c_{n,m})_{n,m\in\Z}$ of a $\torus$-covarian localization observable $E$.
In Section~$\ref{direct}$ we prove by direct methods a
characterization of $\torus$-covariant localization observables in
terms of the matrix $(c_{n,m})_{n,m\in\Z}$.  The same result  can  also 
be obtained from a theorem of Holevo  \cite[Theorem 1]{Holevo87}, whose proof is
also based on group theoretical arguments.

\section{Covariant phase observables}

Theorem \ref{localization} leads also to a characterization of
the covariant phase observables. To describe them, let 
$\mathcal H$ be a complex separable Hilbert space, and let $(|n\rangle)_{n\in\N}$
be an orthonormal basis of $\hi$. We call it the number basis. We define the number operator
$$
N:=\sum_{n\in\N} n|n\rangle\langle n|
$$
with the  domain 
$\mathcal D(N):=\{\psi\in\mathcal H\, : \,
\sum_{n\in\N} n^2|\langle n|\psi\rangle|^2<\infty\},
$
and the unitary `phase shifter' as
$$
U^N(a):= \sum_{n\in\N} a^n|n\rangle\langle n|,
$$
for all $a\in\torus$.
We say that a positive normalized operator measure $\widetilde E:\boto\to\mathcal L(\hi)$
is a {\em phase observable} if it is covariant under the phase shifts, that is, 
if for any $X\in\boto,a\in\torus$,
\begin{equation}
U^N(a)\widetilde E(X)U^N(a)^* = \widetilde E(aX).
\end{equation}

To determine all the phase observables, let $T:\hi\to\lto$ be the linear isometry
with the property
$$
T|n\rangle = e_n,\ {\rm for\ all }\ n\in\N.
$$ 
Clearly, $T$ intertwines the unitary actions
$U^N$  and $U$, $TU^N=UT$, and
$X\mapsto T\widetilde E(X)T^*$
is a $\torus$-covariant localization observable acting in $\lto$. Using 
Theorem \ref{localization}, and the fact that $T^*T=I$, one has the following result.

\begin{corollary}\label{phase}
A normalized positive operator measure $\widetilde E:\boto\to\mathcal L(\hi)$
is a phase observable if and only if it is of the form
$$
\widetilde E(X) = T^*E(X)T,\ X\in\boto,
$$
for some $\torus$-covariant localization observable $E$.
 
Equivalently, $\widetilde E:\boto\to\mathcal L(\hi)$
is a phase observable if and only if
$$
\widetilde E(X) = \sum_{n,m\in\N}\langle \xi_n|\xi_m\rangle
\int_X z^{m-n}\mathrm d\mu(z)\,|n\rangle\langle m|, \ X\in\boto,
$$
for some sequence of unit vectors $(\xi_n)_{n\in\N}$ of $\hi$.
Two sequences of unit vectors $(\xi_n)_{n\in\N}$ and $(\eta_n)_{n\in\N}$ 
define the same phase observable exactly when 
$\langle \xi_n|\xi_m\rangle=\langle \eta_n|\eta_m\rangle$ for all $n,m\in\N$.

Two phase observables $\widetilde E$ and $\widetilde E'$ are equivalent (in the sense of covariance systems)
if and only if any of their generating vector sequences $(\xi_n)$ and $(\xi_n')$ 
are such that, for each $n,m\in\N$, $\langle\xi_n'|\xi_m'\rangle=
\langle z_n\xi_n|z_m\xi_m\rangle$ for some $z_n,z_m\in\torus$. 

Since $T:\hi\to\lto$ is not surjective, there is no projection valued
phase observable.
\end{corollary}

We note, in addition, that Proposition~\ref{commutative}, when applied to phase observables,
gives Eq. (\ref{nmk}) for all $n,m,k\in\N$ with $m\geq k$. For $n=m$ this gives
$|c_{n,n+k}|=|c_{n-k,n}|$ for all $n\geq k$, which implies that $c_{n,m}=0$ for all $n\ne m$
(for details, see \cite{BLPY}). This means that
the only commutative phase observable is the trivial one
$$\boto\ni X\mapsto\mu(X)I\in\mathcal{L(H)}.$$

Following \cite{JPPL99}
we say that a positive semidefinite complex matrix 
$(c_{n,m})_{n,m\in\N}$ is a {\em phase matrix} if $c_{n,n}=1$, for all $n\in\N$.
According to \cite[Phase Theorem 2.2]{JPPL99}, any phase observable $\widetilde E:\boto\to\mathcal L(\hi)$
is of the form  
$$
\widetilde E(X) = \sum_{n,m\in\N} c_{n,m}
\int_X z^{m-n}\mathrm d\mu(z)\, |n\rangle\langle m|
$$
for a unique phase matrix $(c_{n,m})$, and any phase matrix determines a phase observable
in this  way. 
The equivalence of the two characterizations of the phase observables 
is again a consequence of Proposition~\ref{jono}. 

\section{Covariant localizations in a box: a direct method}\label{direct}

We determine next the covariant localizations
by direct methods, using only basic analysis and measure theory. 
Actually, we determine all the normalized (not necessarily positive nor 
self-adjoint) operator measures which are translation covariant on the interval
$[0,2\pi)$. In the rest of this paper, we use 
the interval $[0,2\pi)$ instead of $\torus$ when it simplifies the calculations.
Note that the Haar measure $\mu$ is the normalized Lebesgue measure 
on $\mathcal B([0,2\pi))$, the Borel $\sigma$-algebra of $[0,2\pi)$, 
transferred by the map $\theta\mapsto e^{i\theta}$.
 
Let, again,  $\hi$ be a complex separable Hilbert space, but choose now
an orthonormal basis $(|n\rangle)_{n\in\Z}\subset\hi$ labeled by the integers.
Define the  "extended number operator" as follows:
$\hat N :=\sum_{n\in\Z} n|n\rangle\langle n|$
with its  domain 
$\mathcal D(\hat N):=\{\psi\in\mathcal H\,: \,
\sum_{n\in\Z} n^2|\langle n|\psi\rangle|^2<\infty\}$,
and define the corresponding unitary  shift operators as
$$
R(\theta):=e^{i\theta \hat N}=\sum_{n\in\Z} e^{in\theta}|n\rangle\langle n|
$$
for all $\theta\in\mathbb R$.

We say that $E:\,\mathcal B([0,2\pi))\to\mathcal{L(H)}$ is an operator measure
if it is $\sigma$-additive with respect to the weak operator topology.
If $E(X)^*=E(X)$, or 
$E(X)\ge O$, for all $X\in\mathcal B([0,2\pi))$, we say that $E$ is self-adjoint, or
positive. If $E([0,2\pi))=I$, we say that the operator measure $E$ is normalized.
Finally, $E$ is covariant if $R(\theta)E(X)R(\theta)^*=E(X\oplus\theta)$ for all
$X\in\mathcal B([0,2\pi))$ and $\theta\in\mathbb R$, where the symbol 
$\oplus$ means  addition modulo $2\pi$.

Before characterizing covariant normalized operator measures
 we prove the following lemma:
\begin{lemma}
Fix $q\in\mathbb Z$, and let
$\nu_q:\,\mathcal B([0,2\pi))\to\mathbb C$ be a $\sigma$-additive set function
such that $\nu_q(X\oplus\theta)=e^{iq\theta}\nu_q(X)$ for all $X\in\mathcal B([0,2\pi))$
and $\theta\in[0,2\pi)$, and for which $\nu_q([0,2\pi))=\delta_{0,q}$.
Then $\nu_q(X)=c_q(2\pi)^{-1}\int_Xe^{i q\theta}\mathrm d\theta$ for all
$X\in\mathcal B([0,2\pi))$, where $c_q\in\mathbb C$ and $c_0=1$.
\end{lemma}

\begin{proof}
Fix $q\in\mathbb Z$, and let $k\in\Zp$. Now
\begin{eqnarray}
\delta_{0,q}&=&\nu_q([0,2\pi))=\nu_q\left(\bigcup_{l=0}^{k-1}
\Big[l2\pi k^{-1},(l+1)2\pi k^{-1}\Big)\right)\nonumber\\
&=&\sum_{l=0}^{k-1}\nu_q\Big(\left[0,2\pi k^{-1}\right)+l2\pi k^{-1}\Big)
=\left[\sum_{l=0}^{k-1}e^{i2\pi qk^{-1}l}\right]\nu_q\left[0,2\pi
k^{-1}\right)\nonumber\\
&=&\left\{ \begin{array}{ll}
k\nu_q\left[0,2\pi k^{-1}\right), & \textrm{when $qk^{-1}\in\Z$,}\\
0 & \textrm{when $qk^{-1}\notin\Z.$}
\end{array}\right.\label{2}
\end{eqnarray}

Suppose that $q\in\mathbb Z$ and $k\in\Zp$ are such that $qk^{-1}\notin\Z$. 
Then $\int_0^{2\pi k^{-1}}e^{iq\theta}\mathrm d\theta\ne0$, and we can define
$$
c_q(k):=\frac{\nu_q([0,2\pi k^{-1}))}
{(2\pi)^{-1}\int_0^{2\pi k^{-1}}e^{iq\theta}\mathrm d\theta},
$$
so that
$$
\nu_q([0,2\pi k^{-1}))=c_q(k)
\frac{1}{2\pi}\int_0^{2\pi k^{-1}}e^{iq\theta}\mathrm d\theta=
c_q(k)\frac{e^{iq2\pi k^{-1}}-1}{iq2\pi}.
$$
On the other hand, for all $r\in\Zp$, $q(rk)^{-1}\notin\Z$, and 
\begin{eqnarray*}
\nu_q([0,2\pi k^{-1}))&=&
\nu_q\left(\bigcup_{l=0}^{r-1}
\Big[l2\pi(rk)^{-1},(l+1)2\pi(rk)^{-1}\Big)\right)\\
&=&\left[\sum_{l=0}^{r-1}e^{i2\pi q(rk)^{-1}l}\right]
\nu_q([0,2\pi(rk)^{-1}))\\
&=&c_q(rk)\frac{e^{iq2\pi k^{-1}}-1}{iq2\pi}.
\end{eqnarray*}
This shows that $c_q(k)=c_q(rk)$, $r\in\Zp$. 
Since $q(|q|+1)^{-1}\notin\Z$, one has
$c_q(k)=c_q((|q|+1)k)
=c_q(|q|+1)$. 
Thus,
for all $k\in\Zp$, for which $qk^{-1}\notin\Z$, the number $c_q(k)$ 
is the same, and we may define $c_q:=c_q(|q|+1)$ for all $q\in\Z$
and $q\ne0$.

If $qk^{-1}\in\Z$, $q\in\Z$, $k\in\Zp$, equation (\ref{2}) gives
\begin{equation}\label{2q2}
\nu_q([0,2\pi k^{-1}))=\frac{\delta_{0,q}}{k}
=\frac{1}{2\pi}\int_{0}^{2\pi k^{-1}}e^{iq\theta}\mathrm d\theta.
\end{equation}
Thus, if we define $c_0:=1$ we get  
\begin{equation}\label{b}
\nu_q([0,2\pi k^{-1}))=
c_q\frac{1}{2\pi}\int_0^{2\pi k^{-1}}e^{iq\theta}\mathrm d\theta,
\end{equation}
for all $k\in\Zp$ and $q\in\Z$.

Let $q\in\Z$. Now one gets
$$
\nu_q\left(\bigcup_{p=1}^\infty\left\{p^{-1}\right\}\right)=
\nu_q(\{0\})\sum_{p=1}^\infty e^{iqp^{-1}},
$$
which implies that $\nu_q(\{0\})=0$. 
Thus the measure $\nu_q$ is non-atomic,
that is,
$\nu_q(\{x\})=e^{ixq}\nu_q(\{0\})=0,$ $x\in[0,2\pi),$
which implies that its distribution function $x\mapsto \nu_q([0,x))$ is continuous.
From Equation (\ref{b}) it follows that
for all $k\in\Zp$, $p\in\{1,2...,k\}$, 
\begin{eqnarray}
\nu_q([0,2\pi p k^{-1}))&=&\nu_q\left(\bigcup_{l=0}^{p-1}
\Big[l2\pi k^{-1},(l+1)2\pi k^{-1}\Big)\right)\nonumber\\
&=&c_q\frac{1}{2\pi}\int_0^{2\pi pk^{-1}}e^{iq\theta}\mathrm d\theta.
\end{eqnarray}
Since $x\mapsto \nu_q([0,x))$ is continuous, 
and the set $\big\{2\pi p k^{-1}\in[0,2\pi)\big|
k\in\Zp$, $p\in\{1,2,...,k\}\big\}$ 
is dense in $[0,2\pi)$, it follows that for all $x\in(0,2\pi]$   
$$
\nu_q([0,x))=c_q\frac{1}{2\pi}\int_0^x e^{iq\theta}\mathrm d\theta.
$$
By the Hahn extension theorem 
\begin{equation}\label{a}
\nu_q(X)=c_q\frac{1}{2\pi}\int_X e^{iq\theta}\mathrm d\theta
\end{equation}
for all $X\in\mathcal B([0,2\pi))$ and $q\in\Z$. 
\end{proof}

\begin{theorem}\label{theorem1}
Let $E:\,\mathcal B([0,2\pi))\to\mathcal{L(H)}$ be a covariant normalized 
operator measure. For any $X\in\mathcal B([0,2\pi))$,
\begin{equation}\label{z}
E(X)=  
\sum_{n,m\in\Z} c_{n,m}\frac{1}{2\pi}\int_Xe^{i(n-m)\theta}\mathrm{d}\theta
|n\rangle\langle m|,
\end{equation}
where $c_{n,m}\in\mathbb C$ and $c_{n,n}=1$ 
for all $n,m\in\mathbb Z$. If $E$ is self-adjoint, then 
$\overline{c_{n,m}}=c_{m,n}$ for all $n,m\in\mathbb Z$, and if $E$ is positive, then
\begin{equation}\label{zz}
\sum_{n,m=-k}^k c_{n,m}|n\rangle\langle m|\ge O,
\end{equation}
for all $k\in\mathbb N$.
\end{theorem}

\begin{proof}
 Denoting, in Lemma (1), $q=n-m$ and $\nu_q(X)=\langle n|E(X)|m\rangle$  
Equation (\ref{z}) follows.
If $E$ is self-adjoint, then from (\ref{z}), one gets
$$
\overline{c_{n,m}}=
2\pi\lim_{\epsilon\to0+}\frac{\overline{\langle n|E([0,\epsilon))|m\rangle}}{\epsilon}=
c_{m,n}
$$
for all $n,m\in\Z$. 

Suppose that $E$ is positive and, thus, self-adjoint.
Hence, if (\ref{zz}) does not hold, one may choose 
a $\varphi\in\mathcal H$ and an $l\in\N$ such that
$\sum_{n,m=-l}^lc_{n,m}\langle\varphi|n\rangle\langle m|\varphi\rangle<0$,
and define a function
$$
g:[0,2\pi)\to\mathbb R,\;\theta\mapsto g(\theta):=
\sum_{n,m=-l}^lc_{n,m}e^{i(n-m)\theta}\langle\varphi|n\rangle\langle m|\varphi\rangle.
$$
Due to the continuity of $g$ one can choose an $\epsilon\in(0,2\pi)$ such that
$\int_0^\epsilon g(\theta)\mathrm d\theta<0$. Thus, denoting 
$I_l:=\sum_{n=-l}^l|n\rangle\langle n|$,
$$
\left\langle I_l\varphi|E([0,\epsilon))
I_l\varphi\right\rangle=\frac{1}{2\pi}
\int_0^\epsilon g(\theta)\mathrm d\theta<0,
$$
which contradicts the positivity of $E$. 
\end{proof}

For later use we note that the positive semi-definiteness condition  (\ref{zz}) of the  
matrix  $(c_{n,m})_{n,m\in\mathbb Z}$ 
can be written equivalently as  the following determinant condition, 
see, e.g. \cite[Chpt 3.1]{Bergetal}:
\begin{equation}\label{xx}
\left|
\begin{array}{cccc}
c_{k_1,k_1} & c_{k_1,k_2} & \ldots & c_{k_1,k_s} \\
c_{k_2,k_1} & c_{k_2,k_2} & \ldots & c_{k_2,k_s} \\
\vdots & \vdots & \ddots & \vdots \\
c_{k_s,k_1} & c_{k_s,k_2} & \ldots & c_{k_s,k_s}
\end{array} \right|\ge0
\end{equation}
for all $s\in\Zp$, $\{k_1,\,k_2,...,\,k_s\}\subset\mathbb Z$,
and $k_1<k_2<...<k_s$.
Note that in this case
$c_{n,m}=\overline{c_{m,n}}$ and $|c_{n,m}|\le1$ for all $n,m\in\mathbb Z$.

\begin{remark}\rm
One may ask if the converse statement of Theorem \ref{theorem1} is also true.
Let $(c_{n,m})_{n,m\in\mathbb Z}$ be an infinite-dimensional complex matrix,
and suppose that $c_{n,n}=1$ for all $n\in\mathbb Z$.
Let $\mathcal M:=\mathrm{lin}\{|n\rangle\,|\,n\in\mathbb Z\}$,
and define the following function for all $\varphi$, $\psi\in\mathcal M$:
$$
\mathbb R\ni\theta\mapsto C_{\varphi,\psi}(\theta):=
\sum_{n,m=-\infty}^\infty 
c_{n,m}e^{i(n-m)\theta}\langle\varphi|n\rangle\langle m|\psi\rangle\in\mathbb C.
$$
For $\varphi$, $\psi\in\mathcal M$,  define
$$
E_{\varphi,\psi}([0,2\pi)):=
\frac{1}{2\pi}\int_0^{2\pi}C_{\varphi,\psi}(\theta)\mathrm{d}\theta
=\langle\varphi|\psi\rangle,
$$
Clearly,
$E_{\varphi,\psi}([0,2\pi))=\langle\varphi|\psi\rangle$,
and $(\varphi,\psi)\mapsto E_{\varphi,\psi}([0,2\pi))$ 
is a bounded sesquilinear form 
defined on the dense  subspace $\mathcal M$ of $\mathcal H$. 
Hence the mapping $(\varphi,\psi)
\mapsto E_{\varphi,\psi}([0,2\pi))$
has a unique bounded extension to $\mathcal H$ which is
$(\varphi,\psi)\mapsto\langle\varphi|\psi\rangle$. 
We let $E([0,2\pi))$ denote the unique bounded operator, which, actually,
is the identity operator $I$.

Consider the following sesquilinear form defined for all $X\in\mathcal B([0,2\pi))$:
$$
\mathcal M\times\mathcal M\ni
(\varphi,\psi)\mapsto E_{\varphi,\psi}(X):=
\frac{1}{2\pi}\int_XC_{\varphi,\psi}(\theta)\mathrm{d}\theta\in\mathbb C.
$$
This form 
need not be
bounded, so that it does not necessarily define a 
bounded operator on $\mathcal H$. 
Thus, the formal notation
\begin{equation}\label{gu}
E(X)=\sum_{n,m\in\Z} c_{n,m}\frac{1}{2\pi}\int_X e^{i(n-m)\theta}
\mathrm{d}\theta|n\rangle\langle m|,\;\;\;X\in\mathcal B([0,2\pi)),
\end{equation}
must be understood as the sesquilinear form 
$(\varphi,\psi)\mapsto E_{\varphi,\psi}(X)$ defined on $\mathcal M$.

Since $R(\theta)\mathcal M=\mathcal M$, it follows that
$X\mapsto E(X)$ is covariant in the sense that
$$
E_{R(\theta)^*\varphi,R(\theta)^*\psi}(X)=E_{\varphi,\psi}(X\oplus\theta)
$$
for all $\varphi$, $\psi\in\mathcal M$, $\theta\in[0,2\pi)$, and 
$X\in\mathcal B([0,2\pi))$.

Finally, if $E_{\varphi,\varphi}(X)\ge0$ for all 
$X\in\mathcal B([0,2\pi))$ and $\varphi\in\mathcal M$
we say that $E$ is {positive}. 
If $E$ is positive then
the matrix $(c_{n,m})_{n,m\in\mathbb Z}$ is positive semidefinite
(see the proof of Theorem \ref{theorem1}). Hence,
$0\le E_{\varphi,\varphi}(X)\le
E_{\varphi,\varphi}([0,2\pi))=\|\varphi\|^2$,
$\varphi\in\mathcal M$, $X\in\mathcal B([0,2\pi))$,
and $(\varphi,\psi)\mapsto E_{\varphi,\psi}(X)$ is bounded on $\mathcal M$.
In this case, the sesquilinear form
$E(X)$, for all $X\in\mathcal B([0,2\pi))$, can be regarded as a bounded operator with
the unique matrix elements $E_{|n\rangle,|m\rangle}(X)=c_{n,m}(2\pi)^{-1}\int_X
e^{i(n-m)\theta}\mathrm d\theta$, $n,m\in\mathbb Z$.
The mapping $E:\,\mathcal B([0,2\pi))\to\mathcal{L(H)}$ is $\sigma$-additive (see the 
proof of Phase Theorem 2.2 of \cite{JPPL99}).
Thus, Equation (\ref{gu}) defines weakly a covariant normalized positive operator measure
$X\mapsto E(X)$. 
\end{remark}

\subsection{Projection valued covariant normalized positive operator measures}
The application of \cite[Proposition 1]{Cattaneo} 
in Section~\ref{covloc} gave, in Theorem~\ref{localization}, also 
a characterization of the $\torus$-covariant projection measures. In the present approach 
one has to  determine separately  which of the solutions of Theorem \ref{theorem1}
are projection valued. We shall do that next.

\begin{proposition}
Let $E:\,\bo\to\mathcal{L(H)}$ be a covariant normalized positive operator measure
with the associated structure matrix $(c_{n,m})_{n,m\in\Z}$.
$E$ is projection valued, that is, $E(X)^2=E(X)$ for all $X\in\bo$,
if and only if $|c_{n,m}|=1$ for all $n,\,m\in\Z$.
\end{proposition}

\begin{proof}
Let $x\in(0,1)$ and $n\in\Z$. Using the equations
$\sum_{k=1}^\infty k^{-2}=\pi^2/6$ and 
$y^2=\pi^2/3+4\sum_{k=1}^\infty k^{-2}\cos(k(y+\pi))$, $y\in(-\pi,\pi)$,
one gets
\begin{eqnarray*}
\left\langle n\left|E([0,2\pi x))^2\right|n\right\rangle&=&
\sum_{s=-\infty}^\infty|c_{n,s}|^2\left|\frac{1}{2\pi}\int_0^{2\pi x}e^{i(s-n)\theta}
\mathrm d\theta\right|^2\;\;\;(\hbox{put $k:=s-n$})\\
&\le&x^2+\frac{1}{2\pi^2}\sum_{k=1}^\infty
\frac{\left|e^{2\pi i k x}-1\right|^2}{k^2}\\
&=&x^2+\frac{1}{\pi^2}\sum_{k=1}^\infty\frac{1}{k^2}-\frac{1}{\pi^2}\sum_{k=1}^\infty\frac{\cos(2\pi x k)}{k^2}\\
&=&x=\left\langle n\left|E([0,2\pi x))\right|n\right\rangle
\end{eqnarray*}
where the equality sign holds only when $|c_{n,m}|=1$ for all $n,\,m\in\Z$.

On the other hand, if $|c_{n,m}|=1$, then 
$c_{n,m}=e^{i(\upsilon_n-\upsilon_m)}$, $\upsilon_n\in[0,2\pi)$, 
for all $n,\,m\in\Z$, since $(c_{n,m})_{n,m\in\Z}$ is the structure matrix of $E$  \cite{JPPL00}.
Define the following unitary transformations:
$W:\,\mathcal H\to\mathcal H,\,|n\rangle\mapsto e^{-i\upsilon_n}|n\rangle$
and
$T:\,\mathcal H\to L^2[0,2\pi),\,|n\rangle\mapsto f_n$, where
$f_n(x)=1/\sqrt{2\pi}e^{-inx}$, $x\in[0,2\pi)$.
Now $E$ is unitarily equivalent to 
the canonical spectral measure $E^Q$,
$E^Q(X)f=\chi_Xf$, $X\in\mathcal B([0,2\pi))$, $f\in L^2[0,2\pi)$,
that is, $E(X)=W^* T^* E^Q(X) T W$, and, thus, $E$ is
projection valued. 
\end{proof}

\begin{remark}\rm
The $\torus$-covariant localization observables $E:\boto\to\mathcal{L(H)}$
are compactly supported, ${\rm supp}\, E =\torus$. Therefore, all their moment operators
\begin{eqnarray*}
V^{(k)}& =& \int_\torus z^k\,{\rm d}E(z),\\
E^{(k)} &=& \int_\torus\arg(z)^k\,{\rm d}E(z),
\end{eqnarray*}
with $k\in\Z$, are bounded operators.
The cyclic moments $V^{(k)}$ are contractions whereas  the phase moments $E^{(k)}$ are 
self-adjoint.
The operator measure $E$ is uniquely determined by both of its moment operator sequences
$(V^{(k)})_{k\in\Z}$  and $(E^{(k)})_{k\in\Z}$.

The operator measure $E$ is projection valued if and only if all its cyclic moment operators $V^{(k)}$,
$k\in\Z$, are unitary. If $E$ is not projection valued then, at least, some of the moment operators
$V^{(k)}$ are non-unitary. However, if the first cyclic moment operator $V^{(1)}$ of $E$ is unitary,
then $E$ is projection measure. Indeed,
$$
V^{(1)} =\sum_{n\in\Z} c_{n,n+1}|n\rangle\langle n+1|,
$$
so that $V^{(1)}(V^{(1)})^*=I$ implies that $|c_{n,n+1}|=1$ for all $n\in\Z$. 
By induction, using (\ref{xx}), one then quickly computes that $|c_{n,n+k}|=1$ for all $n\in\Z$
and $k\in\Z^+$, which confirms that $E$ is projection measure (and hence all $V^{(k)}$ are unitary).

We recall further that $E^{(2)}=(E^{(1)})^2$ exactly when $E$ is projection valued 
\cite[Appendix, Sect.\ 3]{FRBSzN}.
In view of that,  it is interesting to observe 
that, due to the covariance condition,
the operator measure  $E$ (projection valued or not)
is uniquely determined already by its first phase moment operator
\begin{eqnarray*}
E^{(1)} &=& \int_\torus\arg(z)\mathrm dE(z) = \int_{0}^{2\pi}\theta\mathrm dE(\theta)\\
&=&
\pi I+\sum_{n\ne m=-\infty}^\infty\frac{c_{n,m}}{i(n-m)}|n\rangle\langle m|
\end{eqnarray*}
since
$c_{n,m}=i(n-m)\langle n|E^{(1)}|m\rangle$ for all $n\ne m$.
Clearly,  the spectral measure $E^{E^{(1)}}$ of the bounded self-adjoint operator $E^{(1)}$
is shift covariant if and only if it is unitarily equivalent to $E^Q$.
\end{remark}



\end{document}